\documentclass[pra,superscriptaddress,showpacs,10pt]{revtex4}
\usepackage{graphicx}
\usepackage{amssymb}

\newtheorem{theorem}{Theorem}
\newtheorem{definition}{Definition}

\newtheorem{proposition}{Proposition}

\begin{document}

\title{Quantum computation based on d-level cluster state }

\author{D.L. Zhou}
\affiliation{Center for Advanced Study, Tsinghua University,
Beijing 100084, China}

\author{B. Zeng}
\affiliation{Department of Physics, Tsinghua University, Beijing,
100084, China}

\author{Z. Xu}
 \affiliation{Center for Advanced Study, Tsinghua
University, Beijing 100084, China} \affiliation{Department of
Physics, Tsinghua University, Beijing, 100084, China}

\author{C.P. Sun}
\affiliation{Institute of Theoretical Physics, The Chinese Academy
of Sciences, Beijing, 100080, China}

%\author{D. L. Zhou$^1$, B. Zeng$^2$, Z. Xu$^{1,2}$ and C. P. Sun$^3$}

%\address{$^1$Center for Advanced Study, Tsinghua University,
%Beijing, 100084, China}
%\address{$^2$Department of Physics, Tsinghua University,
%Beijing, 100084, China}
%\address{$^3$Institute of Theoretical Physics, The Chinese
%Academy of Sciences, Beijing, 100080, China}
\date{\today }

\begin{abstract}
The concept of qudit (a d-level system) cluster state is proposed
by generalizing the qubit cluster state (Phys. Rev. Lett.
\textbf{86}, 910 (2001)) according to the finite dimensional
representations of quantum plane algebra. We demonstrate their
quantum correlations and prove a theorem which guarantees the
availability of the qudit cluster states in quantum computation.
We explicitly construct the network to show the universality of
the one-way computer based on the defined qudit cluster states and
single-qudit measurement. And the corresponding protocol of
implementing one-way quantum computer can be suggested with the
high dimensional ``Ising" model which can be found in many
magnetic systems.
\end{abstract}

\pacs{03.67.Lx}

\maketitle

\section{Introduction}

Quantum computers can process computational tasks that are
intractable with classical computers. The reason lies in the fact
that quantum computing systems composed by qubits (two-level
quantum systems) possess mysterious quantum coherence, such as
entanglement (or quantum correlation), which has no counterpart in
the classical realm \cite{Ni}. Recently, an important kind of
entangled states, cluster states \cite{Bri}, was introduced with
remarkable property---the maximal connectedness, \textit{i.e.},
each pair of qubits can be projected onto maximally entanglement
state with certainty by single-qubit measurements on all the other
qubits. More surprisingly, it was shown that the cluster states
can be used to build a one-way universal quantum computer, in
which all the operations can be implemented by single-qubit
measuments only \cite{Rau}. As one of the key elements to realize
such a scalable quantum computer, a celebrated theorem proved by
R. Raussendorf and H. J. Briegel provides a simple criterion for
the functioning of gate simulations on such quantum computer
\cite{Rau1}. It was pointed out that the protocol of cluster state
computers can be easily realized in practical physical systems
since the creation of the cluster states needs only Ising-type
interactions \cite{Bri}. In fact, it has been demonstrated that
the Hamiltonian with such interactions can be easily found in the
solid state lattice systems with proper spin-spin interactions
\cite{Bri}, even in the system for cold atoms in optical lattices
\cite{Du}.

Theoretically, it is natural to ask whether the concept of cluster
states would have its counterpart in higher dimensional Hilbert
space, \textit{i.e.}, the qudits cases, since most of the
available physical systems can not be treated as a two-level
systems even in an approximate way. The answer to this question is
affirmative. Using a pair of non-commutative operators $X$ and
$Z$, which will be identified with the $d$ dimensional irreducible
representations of Manin's quantum plane algebra (QPA) in Sec. II
\cite{Sun}, we find that the qudit cluster states $|\phi
\rangle_{\mathcal{C}}$ can be defined as a common eigenstate of
the tensor product operators
\begin{equation}
X^{\dag}_a\bigotimes\limits_{b\in {\mathtt{nbgh}}(a)}Z_{b},
\end{equation}
where the lower indexes $a$ and $b$ denote qudit $a$ and qudit $b$
in the cluster, and index $b$ is taken in the neighborhood of
index $a$ depending on the cluster structure. Based on this
definition of qudit cluster states, we further show that we can
build an one way universal quantum computer by explicitly
demonstrating how to construct all single qudit unitary gates and
one imprimitive two-qudit gate.

We organize our paper as follows: We first briefly review the
finite dimensional representations of quantum plane algebra in
Sec. II, which provides the main mathematical tools in this
article. As a non-trivial generalization of the qubit cluster
state, the qudit cluster state is defined according to the quantum
plane algebra, and its essential properties of quantum
correlations are analyzed in Sec. III. In Sec.IV, the one-way
quantum computer based on qudit cluster states is proposed with
crucial supports from the proof of a central theorem. Similar to
that for qubit case \cite{Rau1}, this theorem guarantees the
functioning of gate simulations on those computers. Its proof
depends on the subtle understanding about the quantum correlations
of the qudit cluster states. In Sec. V, the universality of the
qudit cluster state computers is proved through the explicit
construction of all the single- and one two-qudit logic gates.
Finally, we give our conclusion in the last section.

\section{Finite dimensional representaions of quantum plane
algebra}

In this section, we will review the mathematics for finite
dimensional representations of Manin's quantum plane algebra,
which will be the main mathematical tool to describe not only
qudit cluster states but also unitary transformation on the
Hilbert space. The Manin's quantum plane is defined by
\begin{equation}
XZ=qZX, \label{qpc}
\end{equation}%
where $q$ is a complex number. Mathematically, it can be proved
that the associated algebra generated by $Z$, $X$  possesses a
$d-$dimensional irreducible representation only for $q^{d}=1$
\cite{Sun}. In this article, we take $q\equiv q_d\equiv e^{i \frac
{2\pi}{d}}$. This special case is first introduced by Weyl
\cite{Wey} and whose completeness is first proved by Schwinger
\cite{Sch}. Obviously, when $d=1,\,q=1$, $X$ and $Z$ can be
regarded as the ordinary coordinates of $R^{2}$ plane; When
$d=2,\,q=-1$, $X$ and $Z$ can be identified with the Pauli
matrices $\sigma _{x}$ and $\sigma _{z}$. In this sense $Z$ and
$X$ can be regarded as the so-called \textquotedblleft generalized
Pauli operators" \cite{Bar1,San,Dab,Got,Pat,Kni}.

In fact, when $q=q_d$, $Z^{d}$ and $X^{d}$ commutate with the
algebra generators, so they belong to the center of QPA. The
Shur's lemma tells us $Z^{d}$ and $X^{d}$ are constants multiples
of the $d-$dimensional identity matrix, \textit{i.e.}, $Z^{d}=zI$\
and $X^{d}$ $=xI$. In general we can normalize them to the
identity. Since the complex
field $C$ is algebraically closed, there must exist an eigen-state $%
|0\rangle $, which satisfies
\begin{equation}
Z|0\rangle=|0\rangle.
\end{equation}
According to Eq. (\ref{qpc}), we obtain all the eigenvalue
equations for operator $Z$
\begin{equation}
Z|k\rangle=q_d^k|k\rangle, \, (k\in Z_d),
\end{equation}
where $|k\rangle =X^{\dagger k}|0\rangle$. This also implies
\begin{equation}
X|k\rangle =|k-1\rangle. \label{xiz}
\end{equation}%
In the $Z$-diagonal representation, the matrices of $X$ and $Z$
are:
\begin{equation}
Z=\left[
\begin{array}{cccccc}
1 & 0 & 0 & \cdots & 0&0 \\
0 & q_d & 0 & \cdots & 0&0 \\
\vdots & \vdots & \vdots& \ddots  & \vdots & \vdots\\
0 & 0 & 0 & \cdots & q_d^{d-2}&0 \\
0 & 0 & 0 & \cdots & 0& q_d^{d-1}%
\end{array}
\right],
\end{equation}
\begin{equation}
X=\left[
\begin{array}{cccccc}
0 & 1 & 0 & \cdots &0& 0 \\
0 & 0 & 1 & \cdots & 0&0 \\
\vdots & \vdots & \vdots& \ddots  & \vdots & \vdots\\
0 & 0 & 0 & \cdots &1& 0 \\
1 & 0 & 0 & \cdots &0& 0%
\end{array}
\right].
\end{equation}

From Eq. (\ref{xiz}), we have
\begin{equation}
X|x(0)\rangle=|x(0)\rangle,
\end{equation}
where
\begin{equation}
|x(0)\rangle=\frac{1}{\sqrt{d}}\sum\limits_{k=0}^{d-1}|k\rangle.
\end{equation}
Similar to the eigenvalue equation of $Z$, we have
\begin{equation}
X|x(j)\rangle=q_d^j|x(j)\rangle,
\end{equation}
where
\begin{equation}
|x(j)\rangle=Z^j|x(0)\rangle=\frac{1}{\sqrt{d}}\sum\limits_{k=0}^{d-1}q_d^{jk}|k\rangle.
\end{equation}

Corresponding to this representation, we can also define $Z(d)$
algebra as generated by $Z$, $X$. Its all basis elements
\begin{equation}
B=\{Z^{j}X^{k},\, (j, k\in Z_{d})\},\label{oba}
\end{equation}%
are called unitary operator bases in Ref. \cite{Sch}. The general
commutation relations for any two basis elements are
\begin{equation}
X^{j}Z^{k}=q_d^{jk} Z^{k}X^{j}.  \label{unibase}
\end{equation}%

In addition, we can replace the generators $Z$ and $X$ with two
other elements in the basis. First, Let $(m,n)$ be the greatest
common factor of integers $m$ and $n$. Then if
$(m_1,n_1)=1,\;(m_1,n_1\in Z_d)$, we can take
\begin{equation}
\bar{Z}=q_d^{-\frac{d-1}{2}m_1n_1}Z^{m_1}X^{n_1},
\end{equation}
where the factor before $Z^m X^n$ makes $\bar{Z}$ have the same
eigenvalues with $Z$. To maintain Eq. (\ref{qpc}),  we take
\begin{equation}
\bar{X}=q_d^{-\frac{d-1}{2}m_2n_2}Z^{m_2}X^{n_2},
\end{equation}
where $(m_2,n_2)=1,\;(m_2,n_2\in Z_d)$, and $m_1n_2-m_2n_1=1$.
From another viewpoint, $\bar{Z}$ and $\bar{X}$ defines a unitary
transformation $U$
\begin{equation}
\bar{Z}=UZU^{\dagger}, \quad \bar{X}=UXU^{\dagger}.
\end{equation}
By the above definition, it is easy to check that all this kind of
unitary transformations form a group. In fact, it is the so-called
Clifford group, which is a useful concept in universal quantum
computation.

\section{Qudit cluster states in quantum plane}

To generalize the concept of qubit cluster states to qudit cases,
we first restrict ourselves to one dimensional lattices for the
sake of the conceptual simplicity. First, let us recall the
definition of one-dimensional cluster states for $N$ qubits. For a
$N$-site lattice, each qubit is attached to a site. As a novel
multi-qubit entanglement state, the cluster state is written as

\begin{equation}
|\phi \rangle_{\mathcal{C}}
=\frac{1}{2^{N/2}}\bigotimes\limits_{a=1}^{N}(|0\rangle
_{a}+|1\rangle _{a}(\sigma_{z})_{a+1}),\label{cqb1}
\end{equation}%
where $(\sigma _{i})_{a}\,(i=x,y,z)$ are the Pauli matrices
assigned for site $a$ in the lattice, and
$$\sigma_z|s\rangle=(-1)^s |s\rangle, \, (s\in\{0,1\}).$$

Analogy to Eq. (\ref{cqb1}), it is natural to conjuncture that the
qudit cluster state in one dimension as

\begin{equation}
|\phi \rangle_{\mathcal{C}}
=\frac{1}{d^{N/2}}\bigotimes\limits_{a=1}^{N}\left(
\sum\limits_{k=0}^{d-1}|k\rangle
_{a}Z^{k}_{a+1}\right),\label{qdcs}
\end{equation}%
where
\begin{equation}
Z_a |k\rangle_a=q_d^k |k\rangle_a, \,\forall a.
\end{equation}

Now we present one of our main result:

\begin{theorem}
The qudit cluster state in one dimension defined by Eq.
(\ref{qdcs}) is a common eigen-state, with eigen-values being
equal to $1$, of the operators $X^{\dag}_a\bigotimes\limits_{b\in
{\mathtt{nbgh}}(a)}Z_{b}$ , \textit{i.e.},

\begin{equation}
X^{\dag}_a\bigotimes\limits_{b\in
{\mathtt{nbgh}}(a)}Z_{b}|\phi\rangle_{\mathcal{C}} =|\phi
\rangle_{\mathcal{C}}, \label{eqclu}
\end{equation}
where
\begin{equation}
{\mathtt{nbgh}}(a)= \left\{
\begin{array}{ll}
\{2\},& a=1,\\
\{N-1\},& a=N,\\
\{a-1,a+1\},& a\notin \{1,N\}.
\end{array}
\right.
\end{equation}

\end{theorem}

\textbf{Proof}. To prove this theorem, we notice that qudit
cluster state (\ref{qdcs}) can be
constructed in the following procedure. We first prepare a product state%
\[
|+\rangle =\bigotimes\limits_{a=1}^{N}|x(0)\rangle _{a},
\]%
then apply a unitary transformation
\begin{equation}
S=\prod\limits_{b-a=1}S_{ab} \label{clugen}
\end{equation}%
to the state $|+\rangle $. Here $S_{ab}$ is defined by as an
intertwining operator
\begin{equation}
S_{ab}|j\rangle_a |k\rangle_b =q_d^{jk}|j\rangle_a |k\rangle_b.
\end{equation}%
It is easy to prove
\begin{equation}
|\phi \rangle_{\mathcal{C}} =S|+\rangle.
\end{equation}%

Since $X^{\dag}_a|+\rangle =|+\rangle $, it is easy to check
\begin{equation}
SX^{\dag}_aS^{\dag }|\phi \rangle_{\mathcal{C}} =|\phi
\rangle_{\mathcal{C}}.
\end{equation}

The next step is to prove
\[
SX^{\dag}_aS^{\dag }=X^{\dag}_a\bigotimes\limits_{b\in
{\mathtt{nbgh}}(a)}Z_{b}
\]%
To this end, for $a,b,c\in\{1,...,N\}$, we observe that
\begin{equation}
S_{ab}X^{\dag}_aS_{ab}^{\dag }=X^{\dag}_a\otimes Z_{b},
\end{equation}
\begin{equation}
S_{ab}X^{\dag}_bS_{ab}^{\dag }=Z_{a}\otimes X^{\dag}_b,
\end{equation}
\begin{equation}
S_{ab}X^{\dag}_c S_{ab}^{\dag }=X^{\dag}_c,\, \forall c\notin \{
a,b\},
\end{equation}
and
\begin{equation}
S_{ab}Z_{c}S_{ab}^{\dag}=Z_{c},\, \forall c.
\end{equation}

This completes the proof. $\Box$

Although the above proof is restricted to one dimensional cluster,
it is convenient to generalize from one-dimensional qudit cluster
to more complex clusters whether in two or three dimensional
space. In fact, for a general cluster $C$ with one qudit on each
site, the cluster state $|\phi \rangle_{\mathcal{C}}$ is defined
by the following eigen-equations:
\begin{equation}
X^{\dag}_a\bigotimes\limits_{b\in {\mathtt{nbgh}}(a)}Z_{b}|\phi
\rangle_{\mathcal{C}} =|\phi\rangle_{\mathcal{C}}. \label{eqgclu}
\end{equation}%
Formally, the definition of a general cluster is the same as one
dimensional case. Different clusters correspond to different
relations of neighbours.

Now we will discuss the properties of quantum correlations in the
above cluster states under single qudit measurements. For
simplicity, we still restrict ourselves to one dimensional case.

First, let us discuss how to describe a von Neumann measurement
for single qudit. Although the generator $Z$ (or $X$) is not
Hermitian, \textit{i.e.}, the eigenvalues of $Z$ (or $X$) are not
real, the non-degenerate eigenstates of $Z$ (or $X$) can still
form a complete orthogonal basis of $H_{d}$. Therefore they can be
used to define a von Neumann measurement. For example, when we
make a measurement marked by $Z$, we mean that we can obtain
different results corresponding to different eigenstates of $Z$.

Next, we discuss the minimal number of single qudit measurements
needed to destroy all the quantum correlations in qudit cluster
states. For the one dimensional cluster state, we find all quantum
entanglement will be destroyed by measuring $Z_{2a},
\,(a=1,2,\cdots, [N/2])$.

Finally, we will discuss the most remarkable property - maximally
connected - of qudit cluster state, \textit{i.e.}, each pair of
qudits in the cluster can be projected onto maximally entanglement
state with certainty by single-qudit measurement with all the
other qudits. In fact, to project arbitrary two qudits in one
dimensional cluster onto maximally entanglement state, we only
need to measure $X$ for the qudits between them, and $Z$ for all
the other qudits. It is easy to find that two or three dimensional
qudit cluster states is also maximally connected. We only need to
find a one-dimensional path connecting these two qudits and
measure $Z$ for all the other qudits which are not on the path,
which reduces the two or three dimensional problem to one
dimension.

{\vskip 2mm} We now discuss the problem of physical implementation
of our one-way quantum computer. Physically the qudit cluster
state (\ref{qdcs}) can be created by Hamiltonian

\begin{equation}
H=-\hbar g\sum\limits_{(a,b)}N^{(z)}_{a}N^{(z)}_{b}, \,(g>0)
\label{ham}
\end{equation}%
where $(a,b)$ denotes sites $a$ and $b$ are nearest neighbors in
the cluster; and $N^{(z)}$ is defined as
\begin{equation}
N^{(z)}=\sum\limits_{k=0}^{d-1}|k\rangle k\langle k|
\end{equation}%
Then we find the intertwining operator $S$ in Eq. (\ref{clugen})
has the the explicit form
\begin{equation}
S=exp\left( -\frac {i} {\hbar} H t_{\mathcal{C}}\right),
\end{equation}%
where the evolution time $t_{\mathcal{C}}=\frac {2\pi} {dg}$.

To associate with more familiar Hamitonian in physics, let us
define the spin-$\frac{(d-1)}{2}$ operator of $z$ direction
\begin{equation}
s_{z}=N^{(z)}-\frac {d-1} {2}.
\end{equation}%
Then we can rewrite Eq. (\ref{ham}) as
\begin{equation}
H=\frac {d-1} {2}\hbar g \sum\limits_{a}\nu_a {(s_z)}_a-\hbar
g\sum\limits_{(a,b)}{(s_z)}_{a}{(s_{z})}_{b},
\end{equation}%
where $\nu_a$ is the number of nearest neighbors for qudit $a$ in
the cluster. Obviously, the interaction Hamiltonian
\begin{equation}
H_{I}=-\hbar g\sum\limits_{(a,b)}(s_{z})_{a}(s_{z})_{b},
\end{equation}%
which is the ferromagnetic Ising type interaction with
spin-$\frac{d-1}{2}$.

Those kinds of Ising model, other than the usual spin-$1/2$ Ising
model, has been one of the most actively studied systems in
condenced matter and statistical physics due to their rich variety
of critical and multicritical phenomena. For example, the spin -
$1$ Ising model with nearest- neighbor interactions and a
single-ion potential is know as the Blume-Emery-Griffiths (BEG)
model \cite{Blu}, the spin - $3/2$ Ising model was introduced to
explain phase transitions in $DyVO4$ and its phase diagrams were
obtained within the mean-field approximation \cite{Siv}. In
addition, higher spin Ising models can be associated with the
magnetic properties of artificially fabricated superlattices. Such
lattices consist of two or more ferromagnetic materials have been
widely studied over the years, because their physical properties
differ dramatically from simple solids formed from the same
materials. The development of film deposition techniques has
aroused great interest in the synthesis and study of superlattices
in other materials. A number of
experimental\cite{Kwo,Maj,Kre,Cam1} and theoretical
works\cite{Izm,Mat,Hin,Qu,Fis,Gri1,Sy,Sab1} have been devoted to
those directions.

\section{Measurement based quantum computation with qudit}

As discussed above, the qudit cluster states exhibit the same
features in quantum entanglement as that for the qubit cluster
states. Then a question arises naturally: Can these natures of
quantum correlations be available for constructing the universal
quantum computations? We will give an affirmative answer to this
question in the following two sections.

In this section, we further generalize the basic concept of
``single qudit quantum measurement" and the corresponding
measurement based quantum computation (MBQC) on qubit clusters to
that on qudit clusters. Along the line to construct MBQC for the
qubit case, we will formulate the corresponding theorem which
relates unitary transformation to quantum entanglement exhibited
by the qudit cluster states. Quantum computations with qudit
clusters inherit all basic concepts of those with qubit clusters.
They include the basic procedure of simulation of any unitary
gate, the concatenation of gate simulation and the method to deal
with the random measurement results. Here we will give a
$d$-dimensional parallel theorem as generalization of the central
theorem $1$ in Ref. \cite{Rau1}.

Before formulating our central theorem, let us introduce the basic elements
for quantum computing with qudit clusters. The main task of quantum
computing with qudits is to simulate arbitrary quantum gate $g$ defined on $%
n $ qudits Hilbert space. For this purpose, the first step for quantum
computing on qudit clusters is to find out a proper cluster $\mathcal{C}(g)$%
. Then we divide it into three sub-clusters: the input cluster $\mathcal{C}%
_{I}(g)$, the body cluster $\mathcal{C}_{M}(g)$, and the output cluster $%
\mathcal{C}_{O}(g)$. As usual we require that the input and output
clusters have the same rank (i.e. the same number of qudits),
$|\mathcal{C}_{I}(g)|=|\mathcal{C}_{O}(g)|=n$. Then we prepare the
initial state in
\begin{equation}
|\Psi(\mathtt{in})\rangle_{\mathcal{C}(g)}
=|\psi(\mathtt{in})\rangle_{\mathcal{C} _{I(g)}} |+\rangle
_{\mathcal{C }_{M(g)}\cup \mathcal{C}_{O(g)}}. \label{Psi}
\end{equation}

In this first step of quantum computing, we entangle the qudits on
the qudit cluster by using the cluster state generator $S$,
\textit{i.e.},
\begin{equation}
|\Phi(\mathtt{in})\rangle_{\mathcal{C}(g)}
=S|\Psi(\mathtt{in})\rangle_{\mathcal{C}(g)}. \label{Phi}
\end{equation}
This step brings the structure information of the qudit cluster
into our computing process, and thus relates it with the
corresponding qudit cluster state.

The second step is to measure all qudits on the cluster in special
space-time dependent basis according to a given measurement
pattern (MP). The definition of MP is given as follows.
\begin{definition}
A measurement pattern ${\mathcal{M}}_{{\mathcal{C}}}$ on a cluster
${\mathcal{C}}$ is a set of unitary matrixes
\begin{equation}
{\mathcal{M}}_{{\mathcal{C}}}=\left\{ u_{a}Z_{a}u_{a}^{\dagger
}\,|\,\,a\in {\mathcal{C}},\,u_a\in SU(d)\right\} ,
\end{equation}%
which determines the one-qudit measured operators $N^{(u)}_{a}$ on
${\mathcal{C}}$, with the explicit form
\begin{equation}
N^{(u)}_{a}=\sum_{s=0}^{d-1}u_{a}|s\rangle_{a}s\,{_{a}\;\!\!}\langle
s|{u_{a}}^{\dagger}.
\end{equation}
\end{definition}

If this measurement pattern $\mathcal{M}_{\mathcal{C}}$ operates
on the initial state $|\Phi(\mathtt{in})\rangle_{\mathcal{C}(g)}$,
the set of measurement outcomes
\begin{equation}
\{{s}\}_{\mathcal{C}}=\left\{ {s}_{a}\in Z_d\,|\,\,a\in
\mathcal{C}\right\}
\end{equation}%
is obtained. Then, modulo norm factor, the resulting state $|\Psi _{\mathcal{%
M}}\rangle _{\mathcal{C}}$ is given by
\begin{equation}
|{\Phi}_{\mathcal{M}_{\mathcal{C}}}^{\{s\}}\rangle
=P^{\{s\}}_{\mathcal{M}_{\mathcal{C}}}\,
|\Phi(\mathtt{in})\rangle_{\mathcal{C}(g)},
\end{equation}
where the pure state projection
\begin{equation}
P^{\{{s}\}}_{\mathcal{M}_\mathcal{C}}=\bigotimes_{k\in \mathcal{C}%
}u_{k}|s\rangle_{k}\, {_{k}\;\!\!}\langle s|u^{\dagger }_{k}.
\end{equation}%
It is worthy pointing out that we always measure $X$ for the input
qudits and $Z$ for the output, which is independent of gate $g$,
\textit{i.e.},
\begin{eqnarray}
\mathcal{M}_{\mathcal{C}_I(g)}&=&\{X_i, \,i\in
\mathcal{C}_I(g)\},\\
\mathcal{M}_{\mathcal{C}_O(g)}&=&\{Z_i, \,i\in \mathcal{C}_O(g)\}.
\end{eqnarray}
Then we reach the final step, to associate the measurement values
with the result of gate $g$ acting on the initial state.

From the above standard procedure of qudit clusters quantum
computation, we learn that what is crucial for this scheme is to
associate a given gate $g$ with a measurement pattern. Although by
now we have no general optimal operational procedure to do this
for practical problem, the following theorem provides a useful
tool in realizing specific gates on the qudit clusters.

\begin{theorem}\label{coruni}
Suppose that the state $|\psi \rangle
_{{\mathcal{C}}(g)}=P^{\{{s}\}}_{{\mathcal{M}}_{
\mathcal{C}_{M}}(g)}\,|\phi \rangle _{{\mathcal{C}}(g)}$ obeys the
$2n$ eigenvalue equations
\begin{eqnarray}
X_{{\mathcal{C}}_{I}(g),i}\left( UX_{i}U^{\dagger
}\right)_{{\mathcal{C }}_{O}(g)}|\psi \rangle _{{\mathcal{C}}(g)}
& = & q_d^{-\lambda _{x,i}}|\psi
\rangle _{{\mathcal{C}}(g)} \label{check1}\\
{Z^{\dagger }}_{{\mathcal{C}}_{I}(g),i}\left( UZ_{i}U^{\dagger
}\right)_{{\mathcal{C }}_{O}(g)} |\psi \rangle _{{\mathcal{C}}(g)}
& = & q_d^{-\lambda _{z,i}}|\psi \rangle
_{{\mathcal{C}}(g)},\label{check2}%
\end{eqnarray}%
with $\lambda _{x,i},\lambda _{z,i}\in Z_{d}$ and $1\leq i\leq n$.
Then, according to the above standard quantum computing procedure,
we have
\begin{equation}
P^{{\{s\}}}_{{\mathcal{M}}_{\mathcal{C}_{I}(g)}}P^{{\{s\}}}_{\mathcal{M}_{{\mathcal{C}}
_{M}(g)}}\,|\Phi(\mathtt{in})\rangle_{{\mathcal{C}}(g)}
\propto(\prod |s_{i}\rangle )_{{\mathcal{C}}_{I}(g)\cup
{\mathcal{C}} _{M}(g)}{|\psi ({\mathtt{out}})\rangle
}_{{\mathcal{C}}_{O}(g)},\label{out}
\end{equation}
where the input and output state in the simulation of $g$ are related via
\begin{equation}
|\psi ({\mathtt{out}})\rangle =UU_{\Sigma }\,|\psi
({\mathtt{in}})\rangle ,
\end{equation}
where $U_{\Sigma }$ is a byproduct operator given by
\begin{equation}
U_{\Sigma }=\bigotimes_{({\mathcal{C}}_{O}(g)\ni
i)=1}^{n}(Z_{i})^{-\lambda _{x,i}-{s}_{i}}(X_{i})^{\lambda
_{z,i}}.
\end{equation}
\end{theorem}

\textbf{Proof}. Let us begin with the case when
\begin{equation}
|\psi(\mathtt{in})\rangle_{\mathcal{C}_I(g)}=|\{t\}\rangle_{\mathcal{C}_I(g)},
\end{equation}
with
$$\{t\}=\{t_1 t_2 \cdots
t_n\}.$$
To associate with $|\psi\rangle_ {{\mathcal{C}}(g)}$, the
initial input state is written as
 \begin{equation}
 |\psi(\mathtt{in})\rangle_{\mathcal{C}_I(g)}= (\sqrt{d})^n
 P^{\{t\}}_{\mathcal{M}^{\prime}_{\mathcal{C}_I(g)}}|+\rangle_{\mathcal{C}_I(g)},\label{psi}
\end{equation}
where
\begin{equation}
\mathcal{M}^{\prime}_{\mathcal{C}_I(g)}=\{Z_i,\, i\in
\mathcal{C}_I(g)\}.
\end{equation}
  In terms of Eqs. (\ref{Psi},\,\ref{Phi},\,\ref{out},\,\ref{psi}), we have
\begin{equation}
(\prod |s_{i}\rangle )_{{\mathcal{C}}_{I}(g)\cup {\mathcal{C}}
_{M}(g)}{|\psi ({\mathtt{out}})\rangle
}_{{\mathcal{C}}_{O}(g)}\propto
P^{\{s\}}_{\mathcal{M}_{\mathcal{C}_I(g)}}
P^{\{t\}}_{\mathcal{M}^{\prime}_{\mathcal{C}_I(g)}}|\psi\rangle_
{{\mathcal{C}}(g)}.
\end{equation}

To find out the equations for the final state ${|\psi
({\mathtt{out}})\rangle}_{{\mathcal{C}}_{O}(g)}$, we take
$P^{\{s\}}_{\mathcal{M}_{\mathcal{C}_I(g)}}
P^{\{t\}}_{\mathcal{M}^{\prime}_{\mathcal{C}_I(g)}}$ acting on
both sides of Eq. (\ref{check1}) and Eq. (\ref{check2}):
\begin{equation}
\left( UX_{i}U^{\dagger }\right)
_{{\mathcal{C}}_{O}(g)}{|\bar{\psi}
({\mathtt{out}})\rangle}_{{\mathcal{C}}_{O}(g)} =
q_d^{-s_i-\lambda _{x,i}}{|\psi
({\mathtt{out}})\rangle}_{{\mathcal{C}}_{O}(g)},\label{a1'}
\end{equation}
\begin{equation}
\left( UZ_{i}U^{\dagger }\right) _{{\mathcal{C}}_{O}(g)}{|\psi
({\mathtt{out}})\rangle}_{{\mathcal{C}}_{O}(g)} = q_d^{t_i-\lambda
_{z,i}}{|\psi
({\mathtt{out}})\rangle}_{{\mathcal{C}}_{O}(g)},\label{a1}
\end{equation}
where the input state for $|\bar{\psi} ({\mathtt{out}})\rangle$
\begin{equation}
|\bar{\psi} ({\mathtt{in}})\rangle=X_i^{\dagger}|\{t\}\rangle.
\end{equation}

Before drawing a conclusion, we need to check the final state is
not a zero vector. In fact, from Eqs. (\ref{check1}) and
(\ref{check2}), the state $U^{\dagger}|\psi \rangle
_{{\mathcal{C}}(g)}$ is the simutaneous eigen-state of operators
$X_{{\mathcal{C}}_{I}(g),i}X_{{\mathcal{C}}_{O}(g),i}$ and
${Z^\dagger}_{{\mathcal{C}}_{I}(g),i}Z_{{\mathcal{C}}_{O}(g),i}$.
We can directly evaluate this state and find out that it has every
components in $Z$-diagonal representation of the input part.
Consequently the final state is indeed a nonzero vector. So from
Eq. (\ref{a1'}) and Eq. (\ref{a1}), we obtain
\begin{equation}
|\psi(\mathtt{out})\rangle_{{\mathcal{C}}_{O}(g)} =e^{i \eta(t)}
UU_{\Sigma}\,|\{t\} \rangle_{{\mathcal{C}}_{O}(g)}. \label{az}
\end{equation}

To further determine the relation between the output state and the
input state, let us consider the other case when
\begin{equation}
|\psi^\prime (\mathtt{in})\rangle_{{\mathcal
C}_I(g)}=|+\rangle_{\mathcal{C}_I(g)}.
\end{equation}
At this time, the final state
\begin{equation}
|\{s_i\}\rangle_{{{\mathcal{C}}_{I}(g)}\cup
{{\mathcal{C}}_{M}(g)}}\, |\psi^{\prime}
(\mathtt{out})\rangle_{{\mathcal C}_O(g)}=
P^{{\{s\}}}_{{\mathcal{C}}_{I}(g)}({X})|\psi\rangle_
{{\mathcal{C}}(g)}.
\end{equation}
Let $P^{{\{s\}}}_{\mathcal{M}_{{\mathcal{C}}_{I}(g)}}$ apply on
both sides of Eq. (\ref{check1}) and Eq. (\ref{check2}), we obtain
\begin{equation}
\left( UX_iU^{\dagger }\right)_{{\mathcal{C
}}_{O}(g)}|\psi^{\prime} (\mathtt{out})\rangle_{{\mathcal
C}_O(g)}= q_d^{-s_i-\lambda _{x,i}}|\psi^{\prime}
(\mathtt{out})\rangle_{{\mathcal C}_O(g)},
\end{equation}
\begin{equation}
\left( UZ_iU^{\dagger }\right)_{{\mathcal{C
}}_{O}(g)}|\bar{\psi^{\prime}} (\mathtt{out})\rangle_{{\mathcal
C}_O(g)}= q_d^{-\lambda _{z,i}}|\psi^{\prime}
(\mathtt{out})\rangle_{{\mathcal C}_O(g)},
\end{equation}
where the input state for $|\bar{\psi^{\prime}}
(\mathtt{out})\rangle$
\begin{equation}
|\bar{\psi^{\prime}} (\mathtt{in})\rangle=Z^{\dagger}_i|+\rangle.
\end{equation}

 From the above equations, we obtain
\begin{equation}
|\psi^{\prime} (\mathtt{out})\rangle_{{\mathcal C}_O(g)} =e^{i
\chi} UU_{\Sigma}\,|+ \rangle_{{\mathcal{C}}_{O}(g)}. \label{ax}
\end{equation}
Substitute Eq. (\ref{az}) into Eq. (\ref{ax}),
\begin{equation}
|\psi^{\prime} (\mathtt{out})\rangle_{{\mathcal C}_O(g)}
=UU_{\Sigma} d^{-\frac n 2}\sum_{\{t\}}e^{i \eta(t)} \,|\{t\}
\rangle_ {{\mathcal{C}}_{O}(g)}. \label{axz}
\end{equation}
Comparing Eq. (\ref{ax}) to Eq. (\ref{axz}), we obtain
\begin{equation}
e^{i \eta(t)}=e^{i \chi}.
\end{equation}
This completes the proof.$\, \Box$

This theorem tells us that, since the cluster states have
remarkable quantum correlations, which has been discussed in Sec.
II, they play an essential role in the realization of arbitrary
unitary gates. More precisely, as long as one cluster can be used
to process a unitary gate for the cluster state, it will work for
arbitrary input states. Therefore, it is sufficient to check the
conditions for the cluster states, \textit{i.e.}, Eq.
(\ref{check1}) and Eq. (\ref{check2}).  To be emphasized, the
special features of the above theorem arising from qudits is that
it is expressed not only in terms of unitary operators $ X$ and
$Z$ , but also in terms of their conjugates. For $d=2$, it exactly
reduces to the Theorem $1$ of Ref. \cite{Rau1}.

Before using the theorem to construct a specific unitary gate, we
need to explain how to deal with the byproduct part $U_{\Sigma }$.
The basic idea is to move $U_{\Sigma }$ to the front of $U$
according to the commutation relations between $U_{\Sigma }$ and
$U$. To complete this operation, the general strategy is to divide
measurements into several steps, in which the subsequent
measurements depend on the results of previous measurements. In
the next section, we will use specific examples to demonstrate how
to construct all basic elementary gates with the help of this
theorem.

\section{Universality of qudit cluster quantum computation}

It is well known that, for qubit quantum computing network, a
finite collection of one qubit unitary operations and ${CNOT}$
gate is enough to construct any unitary transformation in the
network. This conclusion has been shown to work well for qudit
case \cite{Bry}. More precisely, the collection of all one-qudit
gates and any one imprimitive two-qudit gate is exactly universal
for arbitrary quantum computing, where a primitive two-qudit gate
is such a gate which maps all separate states to separate states.
Therefore, in order to prove universality of quantum computation
with qudit clusters, we only need to construct those basic element
gates, and then to integrate them to realize
arbitrary unitary gate. In this section, based on Theorem \ref%
{coruni} in Sec. III, we will build certain qudit clusters to
simulate the basic elementary gates, \textit{i.e.}, any one-qudit
unitary operation and one imprimitive two-qudit unitary
transformation.

\subsection{Realizations of single qudit unitary transformations}

Let us start with single qudit gates. First, we introduce a
proposition for any unitary transformation in $d$ dimensional
Hilbert space.
\begin{proposition}
Let $\{N_i, \;(i\in Z_{d^2-1})\}$ be a Hermitian basis of the
operator space for $d$ dimensional Hilbert space, then any unitary
transformation $U$ has the form
\begin{equation}
U={q_d}^{\sum_i \alpha_i N_i}=\prod_i{q_d}^{ \beta_i N_i},
\end{equation}
where $\alpha_i$ and $\beta_i$ are real numbers.
\end{proposition}
The first equality is obvious, and  we refer to Ref. \cite{pur}
for the proof of the second equality.

By the above proposition, we can divide any qudit gate into a
product of more simple basic ones, \textit{i.e.}, single parameter
unitary transformations. Now, we need to find $d\times d$
independent $N_i$ to simulate all unitary gates for one qudit. To
use the definition of qudit cluster states, we expect that the
single parameter unitary transformations must have deep relations
with the basis elements of QPA. In fact, we find such a way to
introduce $d\times d$ one parameter unitary transformations. This
originates from that the fact that some of the basis elements of
QPA can be used to define a state basis. We find that all the
unitary transformations that do not change every basis state up to
a phase are defined by the property of multi-values for complex
functions. For example, for operator $Z$, we can define
\begin{equation}
Z^{\beta}(\{m\})=q_d^{\beta N(Z,\{m\})}, \,( \beta\in
\mathcal{R}),
\end{equation}
where
\begin{equation}
N(Z,\{m\})=\sum_{n=0}^{d-1} |n\rangle (n+m_n d)\langle
n|,\,(\forall m_n\in \mathcal{Z}).
\end{equation}
Although the above definitions include infinity unitary
transformations, there are only $d$ independent ones, which can be
used to describe the following type of unitary transformations
\begin{equation}
U_Z (\{\alpha\})|n\rangle =q_d^{ \alpha_n} |n\rangle, \,(\forall
n\in Z_d, \alpha\in\mathcal{R}).
\end{equation}
Obviously, these $d$ independent unitary transformations, also the
corresponding $N(Z,\{m\})$, can take the place of  $\{Z^n, n \in
d\}$ in the unitary basis.

A similar argument can be generalized to $\bar{Z}$, which is
defined in Sec. II. More precisely, for operator $\bar{Z}$, we
define
\begin{equation}
\bar{Z}^{\beta}(\{m\})=q_d^{\beta N(\bar{Z},\{m\})}, \,( \beta\in
\mathcal{R}),
\end{equation}
where
\begin{equation}
N(\bar{Z},\{m\})=\sum_{n=0}^{d-1} |n(\bar{Z})\rangle (n+m_n
d)\langle n(\bar{Z})|,\,(\forall m_n\in \mathcal{Z}),
\end{equation}
with
\begin{equation}
\bar{Z}|n(\bar{Z})\rangle=q_d^n|n(\bar{Z})\rangle.
\end{equation}

In the following, we will show that we can select $d\otimes d$
independent Hermitian operators from  $N(\bar{Z},\{m\})$.

When $d$ is a prime number, a convenient choice is to take
$\bar{Z}$ from the operator set
\begin{equation}
\{Z, X, ZX, ZX^2,\cdots, ZX^{d-1}\}.
\end{equation}
Because each $\bar{Z}$ defines $d-1$ independent
$N(\bar{Z},\{m\})$ besides the identity, we obtain
$d^2\,(=(d-1)(d+1)+1)$ independent Hermitian operators.

When $d$ is not a prime number, we can choose the independent
$N(\bar{Z},\{m\})$
 by the following procedure. First, we take $\bar{Z}$ from $\{Z, \, X,\,
ZX\}$, and we can obtain $3(d-1)$ independent $N(\bar{Z},\{m\})$
besides the identity, which can take the place of the set of basis
elements
\begin{equation}
S=\{ Z^n, X^n, Z^nX^n, \,(n\in Z_d)\}.
\end{equation}
Then we take an element $\bar{Z}\notin S$, find out the elements
in $\{\bar{Z}^n, \, n\in Z_d\}$ that is not in the set $S$, add
these elements into $S$, and take the new independent
$N(\bar{Z},\{m\})$, whose number is the number of new elements in
set $S$. We repeat the above step until set $S=B$, then we obtain
$d^2$ independent Hermitian operators.

Therefore if we can do all the above basic unitary transformations
$\bar{Z}^{\beta}(\{m\}$, we will declare that we can do all single
qudit unitary transformations. Here we adopt the following
strategy: First, we realize $Z^{\alpha}(\{m\}$ on a five qudit
cluster, as a basic single qudit transformation. To associate with
all the other single qudit unitary transforamtion
$\bar{Z}^{\alpha}_{\{m\}}$, we observe an useful fact
\begin{equation}
\bar{Z}^{\alpha}(\{m\})=U
Z^{\alpha}(\{m\}){U}^{\dagger},\label{clif}
\end{equation}
where $U$ (or $U^{\dagger}$) can be taken as an element in the
Clifford group as defined in Sec. II. Based on this observation,
we further discuss how to implement these elements in the Clifford
group on clusters. Then we only need to connect these  clusters in
the given order to realized unitary gate
$\bar{Z}^{\alpha}(\{m\})$. If we succeed to pass the above
procedure, in principle we can make any single qudit unitary gate
in $SU(d)$.

Now, let us realize these basic unitary transformations with a
specifically-designed qudit cluster, which will be given as
follows.

\subsubsection{\protect\bigskip Five qudit cluster realization of
$X^\alpha(\{m\})$ and $Z^\alpha(\{m\})$ }

\begin{figure}[htbp]
\begin{center}
\includegraphics{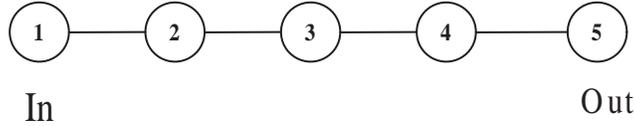}
\end{center}
\caption{Five qudit cluster used in realization of
$X^\alpha(\{m\})$ and $Z^\alpha(\{m\})$. A circle represents one
qudit, number $n$ in the circle means the $n$-th qudit, $in$ or
$out$ denote the input or output part of the cluster, and two
qudits which are connected by a line is neighbors. } \label{fig5}
\end{figure}

In this subsection, we aim to realize the basic single qudit
unitary transformation $Z^\alpha(\{m\})$ on a five qudit cluster
designed as in Fig. \ref{fig5}, which is a linear array of five
qudits. By the way, we also use the same cluster to implement
$X^\alpha(\{m\})$, which demonstrates that the same cluster with
different measurement patterns can realize different unitary
transformations.
The corresponding cluster state is defined by the following system
of equations:
\begin{eqnarray}
X_{1}^{\dagger }Z_{2}|\phi \rangle_{\mathcal{C}} &=&|\phi
\rangle_{\mathcal{C}} , \label{5qdit1}
\\
Z_{1}X_{2}^{\dagger }Z_{3}|\phi \rangle_{\mathcal{C}} &=&|\phi
\rangle_{\mathcal{C}},
\label{5qdit2} \\
Z_{2}X_{3}^{\dagger }Z_{4}|\phi\rangle_{\mathcal{C}} &=&|\phi
\rangle_{\mathcal{C}} ,
\label{5qdit3} \\
Z_{3}X_{4}^{\dagger }Z_{5}|\phi \rangle_{\mathcal{C}} &=&|\phi
\rangle_{\mathcal{C}} ,
\label{5qdit4} \\
Z_{4}X_{5}^{\dagger }|\phi \rangle_{\mathcal{C}} &=&|\phi
\rangle_{\mathcal{C}} . \label{5qdit5}
\end{eqnarray}%
It follows from Eqs. (\ref{5qdit1}-\ref{5qdit5}) that
\begin{eqnarray}
X_{1}X_{3}^{\dagger }X_{5}|\phi \rangle_{\mathcal{C}} &=&|\phi \rangle_{\mathcal{C}} , \\
Z_{1}^{\dagger }X_{2}X_{4}^{\dagger }
Z_{5}|\phi\rangle_{\mathcal{C}} &=&|\phi \rangle_{\mathcal{C}} .
\label{}
\end{eqnarray}%
Form Eq. (\ref{5qdit5}), we obtain
\begin{equation}
Z_4^{\dagger \alpha}(\{m\}) X_5^{\alpha}(\{m\})|\phi
\rangle_{\mathcal{C}}=|\phi \rangle_{\mathcal{C}}.
\end{equation}
Here we emphasizes that when we obtain the above equation, we have
used the following condition On $\{m\}$. If $n_4+n_5=0
\;(Mod(d))$, then
\begin{equation}
n_4+m_{n_4}+n_5+m_{n_5}=0,\; (n_4, n_5\in Z_d).
\end{equation}

From the above four equations, we have
\begin{eqnarray}
X_{1}X_{3}^{\dagger
}\left(X_5^{\alpha}\left(\{m\}\right)X_{5}X_5^{\dagger
\alpha}\left(\{m\}\right)\right)
|\phi \rangle_{\mathcal{C}} &=&|\phi \rangle_{\mathcal{C}} , \\
Z_{1}^{\dagger }X_{2}\left(Z_4^{\dagger
\alpha}(\{m\})X_{4}^{\dagger
}Z_4^{\alpha}(\{m\})\right)\left(X_5^{\alpha}(\{m\})Z_{5}X_5^{\dagger
\alpha}(\{m\})\right)|\phi \rangle_{\mathcal{C}} &=&|\phi
\rangle_{\mathcal{C}} . \label{x20}
\end{eqnarray}%
For a measurement pattern $\{X_{2}, \,X_{3}^{\dagger },\,
Z_4^{\dagger \alpha}(\{m\})X_{4}^{\dagger }Z_4^{\alpha}(\{m\})\}$,
Theorem \ref{coruni} concludes that the simulated unitary
transformation is $X_5^{\alpha}(\{m\})U_{\Sigma }$, where
$$
U_{\Sigma}=Z_5^{-s_1-s_3}X_5^{s_2+s_4}.
$$
 Because
$U_{\Sigma }$ depends on the measurement results and can not be
moved to the front of $X_5^{\alpha}(\{m\})$ trivially, different
measurement results lead to different unitary transformations. In
order to realize unique gate $X_5^{\alpha}(\{m\})$, we complete
the measurement in two steps: In the first step, we measure
$\{X_{1},X_{2},X_{3}^{\dagger }\}$. Then, when the obtained values
are $s_{1}$ , $\lambda _{x}\,(=s_{3})$ and $s_{2}$, the byproduct
operator reads $U_{\Sigma }=Z^{-s_{3}-s_{1}}X^{\lambda _{z}}$. At
this time, $\lambda _{z}$ is still unknown since it depends on the
measured result $s_{4}$. However, notice that $
Z_{5}^{-s_{3}-s_{1}}X_5^{\alpha}(\{m\})Z_{5}^{s_{3}+s_{3}}$ is
diagonal in the $X$ representation, we have
\begin{equation}
Z_{5}^{-s_{3}-s_{1}}X_5^{\alpha}(\{m\})Z_{5}^{s_{3}+s_{1}}=\prod_{\{m\}}
X_5^{\alpha^{s_1 s_3}_{\{m\}}}(\{m\}). \label{sepx}
\end{equation}
From Eq. (\ref{sepx}), we can obtain $d$ equations of
$\{\alpha^{s_1 s_3}_{\{m\}}\}$,
which determine the values of $%
\{\alpha^{s_1 s_3}_{\{m\}}\}$.

Now we make a new choice depending on the known measurement
results. Also from Eqs. (\ref{5qdit1}-\ref{5qdit5}), we obtain the
following equation instead of Eq. (\ref{x20})
\begin{eqnarray}
&&Z_{1}^{\dagger }X_{2}\left(\prod_{\{m\}} Z_4^{\dagger
\alpha^{s_1 s_3}_{\{m\}}}(\{m\})X_{4}^{\dagger }\prod_{\{m\}}
Z_4^{\alpha^{s_1 s_3}_{\{m\}}}(\{m\})\right)  \nonumber \\
&&\left(\prod_{\{m\}} X_5^{\alpha^{s_1
s_3}_{\{m\}}}(\{m\})Z_{5}\prod_{\{m\}} X_5^{\dagger \alpha^{s_1
s_3}_{\{m\}}}(\{m\})\right)|\phi \rangle_{\mathcal{C}} =|\phi
\rangle_{\mathcal{C}} . \label{x201}
\end{eqnarray}%
Measuring the fourth qudit in the base
\[
\prod_{\{m\}} Z_4^{\dagger \alpha^{s_1
s_3}_{\{m\}}}(\{m\})X_{4}^{\dagger }\prod_{\{m\}} Z_4^{\alpha^{s_1
s_3}_{\{m\}}}(\{m\}),
\]%
we obtain the value $s_{4}$ and therefore $\lambda
_{z}=s_{2}+s_{4}$. According to Theorem \ref{coruni}, we obtain
the final operation
\begin{equation}
UU_{\Sigma }=Z_{5}^{-s_{3}-s_{1}}X_5^{s_2+s_4}
X_5^{\alpha}(\{m\}).
\end{equation}%
Finally, by measuring $Z_{5}$ we obtain the correct result
\begin{equation}
s=s_{5}+s_{2}+s_{4}.  \label{corres}
\end{equation}

The above equation does not mean that the final result depends
only on the measurements to the second, fourth, and fifth qudit,
because different values of $s_{1}$ and $s_{3}$ correspond to
different measurement on the fourth qudit.

Based on the same cluster, we can also implement the single qudit rotation $%
Z^\alpha (\{m\})$. Similarly, we first measure
$\{X_{1},X_{2},x^{\dagger}_4\}$; Then from Eqs.
(\ref{5qdit1}-\ref{5qdit5}), we obtain
\begin{eqnarray}
X_{1}\left(\prod_{m}Z_3^{\alpha^{s_2 s_4}_{\{m\}}}
(\{m\})X_{3}^{\dagger
}\prod_{m}Z_3^{\dagger \alpha^{s_2 s_4}_{\{m\}}} (\{m\})\right) &&  \nonumber \\
\left(\prod_{m}Z_5^{\alpha^{s_2 s_4}_{\{m\}}}
(\{m\})X_{5}\prod_{m}Z_5^{\dagger \alpha^{s_2 s_4}_{\{m\}}}
(\{m\})\right)|\phi
\rangle_{\mathcal{C}} &=&|\phi \rangle _{\mathcal{C}}, \\
Z_1^{\dagger }X_{2}X_{4}^{\dagger }Z_{5}|\phi
\rangle_{\mathcal{C}} &=&|\phi \rangle_{\mathcal{C}} .
\end{eqnarray}
where $\{\alpha^{s_2 s_4}_{\{m\}}\}$ is determined by
\begin{equation}
X_{5}^{s_{2}+s_{4}}Z_5^{\alpha}({\{m\}})X_{5}^{-s_{2}-s_{4}}=\prod_{\{m\}}Z_5^{\alpha_{\{m\}}^{s_2
s_4}}({\{m\}}).
\end{equation}
At the same time we make another measurement on
$$\prod_{m}Z_3^{\alpha^{s_2 s_4}_{\{m\}}} (\{m\})X_{3}^{\dagger
}\prod_{m}Z_3^{\dagger \alpha^{s_2 s_4}_{\{m\}}} (\{m\}).$$
According to theorem \ref{coruni}, we conclude that the simulated
unitary transformation indeed is
\[
Z^{-s_{3}-s_{1}}X^{s_{2}+s_{4}}Z_5^{\alpha}({\{m\}}).
\]
The correct result and the measurement values are also related by Eq. (\ref%
{corres}).

\subsubsection{Realizations of single qudit elements in Clifford
group}

As implied in Eq. (\ref{clif}), we only need to realize the single
qudit elements in the Clifford group. It is easy to show that not
all elements in the Clifford group are required. In fact, we only
need do the elements defined as
\begin{equation}
U^{mn}Z{U^{mn}}^\dagger=\bar{Z},\label{dmn}
\end{equation}
where
\begin{equation}
\bar{Z}=q_d^{-\frac{d-1} {2} mn}Z^mX^n,
\end{equation}
with
\begin{equation}
(m,n)=1.
\end{equation}

We will show that we can do all the above Clifford unitary
transformations through a series of four basic types of unitary
transformations. The first is defined as
\begin{eqnarray}
U^{1n}Z{{U^{1n}}^\dagger}&=&q_d^{-\frac{d-1} {2} n}Z X^{n},\\
U^{1n}X{{U^{1n}}^\dagger}&=&X;
\end{eqnarray}
The second is defined as
\begin{eqnarray}
U^{n1}Z{{U^{n1}}^\dagger}&=&q_d^{-\frac{d-1} {2} n}Z^{n} X,\\
U^{n1}X{{U^{n1}}^\dagger}&=&Z^{\dagger};
\end{eqnarray}
The third is defined as
\begin{eqnarray}
V Z{{V}^\dagger}&=&q_d^{-\frac{d-1} {2}}Z X,\\
V X{{V}^\dagger}&=&X;
\end{eqnarray}
The last is defined as
\begin{eqnarray}
W Z{{W}^\dagger}&=&Z,\\
W X{{W}^\dagger}&=&q_d^{-\frac{d-1} {2}}Z X.
\end{eqnarray}

\begin{theorem}
For any unitary transformation $U^{mn}$ partially defined by Eq.
(\ref{dmn}), it can be factorized into a product of a series of
the above four basic unitary transformations.
\end{theorem}

\textbf{Proof}. Let us prove it by induction. We denote
$N_d\equiv\{1,2,\cdots, d\}$. When $m=1$ or $n=1$, it is the first
or the second types of unitary transformations. Let us suppose the
above theorem is valid at $m\in N_d$ or $n\in N_d$, \textit{i.e.},
we can do
\begin{equation}
U^{mn}\qquad (m \;\mbox{or}\; n\in N_d,\, (m,n)=1 ).
\end{equation}
For arbitrary positive integer $n$ that is relatively prime to
$d+1$, there exists
\begin{equation}
n=i(d+1)+n^{\prime}, \quad (i\in Z_{\infty},\, n^{\prime}\in N_d).
\end{equation}
Obviously,
\begin{equation}
n^{\prime}\in N_d,\, (d+1,n^{\prime})=1.
\end{equation}
According to the assumption, we can do $U^{(d+1) n^{\prime}}$.
Then let $V$ operate $i$ times,
\begin{equation}
V^i U^{(d+1)n^{\prime}} {V^{\dagger}}^i=U^{(d+1)n}.
\end{equation}
According to the assumption, we can also do $U^{n^{\prime}(d+1)
}$. Then let $W$ operate $i$ times,
\begin{equation}
W^i U^{n^{\prime}(d+1)} {W^{\dagger}}^i=U^{n(d+1)}.
\end{equation}
Therefore the theorem is valid for $m \;\mbox{or}\; n \in
N_{d+1}$. This completes the proof.  $\Box$

Now we come to constructing the four basic unitary
transformations. We will prove that the first (including the
third) and the fourth can be realized on the five qudit cluster as
Fig. {\ref{fig5}}. According to Eqs. (\ref{5qdit1}-\ref{5qdit5}),
we have
\begin{eqnarray}
X_{1}X_{3}^{\dagger }X_{5}|\phi \rangle_{\mathcal{C}} &=&|\phi \rangle_{\mathcal{C}} , \\
Z_{1}^{\dagger }X_{2}({Z_{4}}^n X_{4})^{\dagger }
Z_{5}{X_{5}}^n|\phi\rangle_{\mathcal{C}} &=&|\phi
\rangle_{\mathcal{C}} . \label{}
\end{eqnarray}
According to theorem \ref{coruni}, when the measurement pattern is
$\{X_{2},\,X_{3}^{\dagger },\,({q_d}^{-\frac {d-1} {2} n}{Z_{4}}^n
X_{4})^{\dagger }\}$, the corresponding unitary transformation is
\begin{equation}
{q_d}^{\frac {d-1} {2}
(s_1+s_3)n}Z^{-s_1-s_3}X^{s_2+s_4-n(s_1+s_3)} U^{1n}.
\end{equation}

Also from Eqs. (\ref{5qdit1}-\ref{5qdit5}), we obtain
\begin{eqnarray}
X_{1}Z_{3}X_{3}^{\dagger }X_{4}^{\dagger}Z_{5}X_{5}|\phi \rangle_{\mathcal{C}} &=&|\phi \rangle_{\mathcal{C}} , \\
Z_{1}^{\dagger }X_{2}X_{4}^{\dagger }
Z_{5}|\phi\rangle_{\mathcal{C}} &=&|\phi \rangle_{\mathcal{C}} .
\label{}
\end{eqnarray}
According to theorem \ref{coruni}, when the measurement pattern is
$\{X_{2},\,{q_d}^{\frac {d-1} {2} n}Z_{3}X_{3}^{\dagger },\,
X_{4}^{\dagger }\}$, the corresponding unitary transformation is
\begin{equation}
{q_d}^{\frac {d-1} {2} (s_2+s_4)}Z^{s_2-s_1-s_3}X^{s_2+s_4} W.
\end{equation}

\begin{figure}[htbp]
\begin{center}
\includegraphics{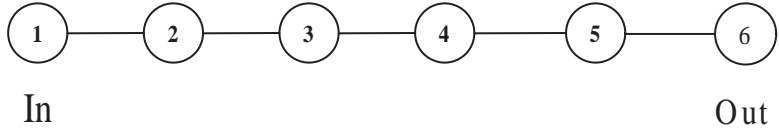}
\end{center}
\caption{Six qudit cluster used in realization of $U^{n1}$.
Meanings of the symbols in this Figure are the same as Fig.
\ref{fig5}.} \label{fig6a}
\end{figure}

To realize the unitary transformation $U^{n1}$, we need a cluster
composed of six qudits as Fig. \ref{fig6a}.
The cluster state is defined by the following system of equations
\begin{eqnarray}
X_{1}^{\dagger }Z_{2}|\phi _{\mathcal{C}}\rangle &=&|\phi
\rangle_{\mathcal{C}} , \label{6qdita1}
\\
Z_1 X_{2}^{\dagger }Z_{3}|\phi\rangle_{\mathcal{C}} &=&|\phi
\rangle_{\mathcal{C}} , \label{6qdita2}
\\
Z_{2}X_{3}^{\dagger }Z_{4}|\phi\rangle_{\mathcal{C}} &=&|\phi
\rangle_{\mathcal{C}} ,
\label{6qdita3} \\
Z_{3}X_{4}^{\dagger }Z_{5}|\phi\rangle_{\mathcal{C}} &=&|\phi
\rangle_{\mathcal{C}},
\label{6qdita4} \\
Z_{4}X_{5}^{\dagger }Z_{6}|\phi \rangle_{\mathcal{C}} &=&|\phi
\rangle_{\mathcal{C}} , \label{6qdita5}
\\
Z_{5}X_{6}^{\dagger }|\phi\rangle_{\mathcal{C}} &=&|\phi
\rangle_{\mathcal{C}} . \label{6qdita6}
\end{eqnarray}%
It follows from the above equations that
\begin{eqnarray}
X_{1}X_{3}^{\dagger}X_{5} Z_{6}^{\dagger}|\phi \rangle_{\mathcal{C}} &=&|\phi\rangle_{\mathcal{C}} , \\
Z_{1}^{\dagger}X_{2}{Z_{4}}^nX_{4}^{\dagger}{X_{5}^{\dagger}}^n{Z_{6}}^nX_{6}|\phi
\rangle_{\mathcal{C}} &=&|\phi \rangle_{\mathcal{C}}.
\end{eqnarray}
When the measurement pattern is $%
\{X_{2},\,X_{3}^{\dagger},\,{q_d}^{\frac {d-1} {2}
n}{Z_{4}}^nX_{4}^{\dagger},\,X_{5}\}$, the corresponding unitary
transformation is
\begin{equation}
{q_d}^{(s_1+s_3)(s_2+s_4-ns_5+\frac {d-1} {2}
n)}Z^{n(s_5-s_1-s_3)-s_2-s_4}X^{-s_1-s_3} U^{n1}.
\end{equation}

\subsection{\protect\bigskip Realization of an imprimitive two qudit gate}

Now we come to the construction for simulating two qudit
operations.
\begin{figure}[htbp]
\begin{center}
\includegraphics{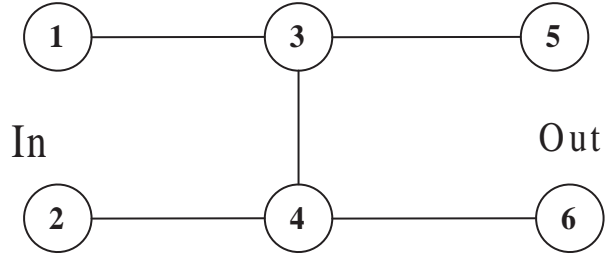}
\end{center}
\caption{Six qudit cluster used in realization of an imprimitive
two qudit gate $T$. Meanings of the symbols in this Figure are the
same as Fig. \ref{fig5}.} \label{fig6}
\end{figure}
The cluster composed of six qudits as Fig. \ref{fig6} is
considered with the following system of equations
\begin{eqnarray}
X_{1}^{\dagger }Z_{3}|\phi _{C}\rangle &=&|\phi
\rangle_{\mathcal{C}} , \label{6qdit1}
\\
X_{2}^{\dagger }Z_{4}|\phi\rangle_{\mathcal{C}} &=&|\phi
\rangle_{\mathcal{C}} , \label{6qdit2}
\\
Z_{1}X_{3}^{\dagger }Z_{4}Z_{5}|\phi\rangle_{\mathcal{C}} &=&|\phi
\rangle_{\mathcal{C}} ,
\label{6qdit3} \\
Z_{2}Z_{3}X_{4}^{\dagger }Z_{6}|\phi\rangle_{\mathcal{C}} &=&|\phi
\rangle_{\mathcal{C}},
\label{6qdit4} \\
Z_{3}X_{5}^{\dagger }|\phi \rangle_{\mathcal{C}} &=&|\phi
\rangle_{\mathcal{C}} , \label{6qdit5}
\\
Z_{4}X_{6}^{\dagger }|\phi\rangle_{\mathcal{C}} &=&|\phi
\rangle_{\mathcal{C}} . \label{6qdit6}
\end{eqnarray}%
It follows from the above equations that
\begin{eqnarray}
X_{1}X_{5}^{\dagger }|\phi \rangle_{\mathcal{C}} &=&|\phi\rangle_{\mathcal{C}} , \\
X_{2}X_{6}^{\dagger }|\phi\rangle_{\mathcal{C}} &=&|\phi \rangle_{\mathcal{C}} , \\
Z_{1}^{\dagger }X_{3}Z_{5}^{\dagger}X_{6}^{\dagger }|\phi
\rangle_{\mathcal{C}} &=&|\phi
\rangle_{\mathcal{C}} , \\
Z_{2}^{\dagger }X_{4}Z_{6}^{\dagger}X_{5}^{\dagger }|\phi
\rangle_{\mathcal{C}} &=&|\phi\rangle_{\mathcal{C}}.
\end{eqnarray}

By measuring the system according to the measurement pattern $%
\{X_{1}X_{2}X_{3}X_{4}Z_{5}Z_{6}\}$, the simulated two qudit gate $T$
satisfies, and is also defined by
\begin{eqnarray}
TX_{5}T^{\dagger } &=&X_{5}^{\dagger }, \\
TX_{6}T^{\dagger } &=&X_{6}^{\dagger }, \\
TZ_{5}T^{\dagger } &=&Z_{5}^{\dagger} X_{6}^{\dagger }, \\
TZ_{6}T^{\dagger } &=&Z_{6}^{\dagger} X_{5}^{\dagger }.
\end{eqnarray}%
According to Theorem $\ref{coruni}$, the above measurement pattern
realize the following unitary gate
\begin{equation}
TZ_{5}^{-s_{1}}X_{5}^{s_{3}}Z_{6}^{-s_{2}}X_{6}^{s_{4}}={q_d}^{
s_{1}s_{2}}Z_{5}^{s_{1}}X_{5}^{s_{2}-s_{3}}Z_{6}^{s_{2}}X_{6}^{s_{1}-s_{4}}T.
\end{equation}

The next task is to prove that $T$ is an imprimitive two qudit
operation. Ref. \cite{Bry} tells us that a two gate $V$ is
primitive if and only if $V=S_{1}\otimes S_{2}$ or
$V=(S_{1}\otimes S_{2})P$. Here, $S_{1}$ and $S_{2}$ are different
single qudit operators, $P$ is the interchanging
operator obeying $P|x\rangle \otimes |y\rangle =|y\rangle \otimes |x\rangle $%
. Based on this fact, we can easily conclude that a primitive
operator always maps a single qudit operator to another single
qudit operator. Obviously, the above two qudit operator $T$ is
imprimitive. Another way to prove it is to evaluate the unitary
transformation directly. Then we can find that it maps all
$Z\otimes Z-$ bases to the maximum entangle states.

As demonstrated in this section, any single qudit unitary gate and
one imprimitive two qudit gate can be realized on qudit clusters.
Therefore, the measurement based quantum computing on qudit
clusters is universal.

\section{Conclusions}

We have introduced the concept of qudit cluster state in terms of
finite dimensional representations of QPA. Based on these qudit
cluster states, we have built all the elements of qudit clusters
needed for implementation of universal measurement-based quantum
computations. With generalizations of cluster states and
measurement patterns, most of the results in qubit cluster can
work well for qudit clusters in parallel ways. We also show that
there still exists the celebrated theorem guaranteeing the
availability of qudit cluster states to quantum computations. To
prove the universality of this quantum computation, we show that
we can implement all single qudit unitary transformations and one
imprimitive two qudit gate on specific qudit clusters. In
addition, we propose to build a one-way universal quantum computer
with qudit cluster states practically since the high dimensional
\textquotedblleft Ising\textquotedblright model can be used to
generate such cluster state dynamically.

\begin{acknowledgements}

The authors would like to thank Prof. X. F. Liu for useful
discussions. The work of D. L. Z is partially supported by the
National Science Foundation of China (CNSF) grant No. 10205022.
The work of Z. X is supported by CNSF (Grant No. 90103004,
10247002). The work of C. P. S is supported by the CNSF( grant No.
10205021)and the knowledged Innovation Program (KIP) of the
Chinese Academy of Science. It is also funded by the National
Fundamental Research Program of China with No 001GB309310.

\end{acknowledgements}

\end{document}